
\font\ninerm=cmr9  
\font\twerm=cmr12
\font\titlefnt=cmbx10 scaled \magstep1
\font\tit=cmbx10 scaled \magstep2
\magnification 1200
\parindent 0pt

\def\pmb#1{\setbox0=\hbox{$#1$}%
\kern-.025em\copy0\kern-\wd0
\kern.05em\copy0\kern-\wd0
\kern-.025em\raise.0433em\box0 }

\rightline{\titlefnt hep-ph/9502342}
\rightline{ DTP/95/16}
\rightline{February 1995}
\null\vskip 1truecm

\centerline{\tit Fermion Number Violation in the  Background}
\smallskip
\centerline{\tit of a Gauge Field in Minkowski Space}
\vskip 2truecm
\centerline{\twerm Valentin V. Khoze \ \footnote{$^\spadesuit$}{\ninerm
E-mail: valya.khoze@durham.ac.uk}}
\vskip .5truecm
\centerline{\it Department of Physics, Science Laboratories}
\smallskip
\centerline{\it University of Durham, Durham, DH1 3LE, United Kingdom}
\vskip 2truecm
\centerline{\it Abstract}
\smallskip
Anomalous fermion number violation is studied in the background of a pure
$SU(2)$ gauge field in Minkowski space using the method of N. Christ.
It is demonstrated that the chiral fermion number is violated by at most
an integer amount. Then the method is applied for a  spherically symmetric
Minkowski space classical gauge field in the background. These classical
gauge fields are finite energy solutions to pure $SU(2)$ equations of motion
with in general non-integer topological charge. We show that
in the classical background which during a finite time-interval matches such
solutions the fermion number violation is integer and non-zero. In particular,
we calculate the violation of the fermion number in the presence of
L\"uscher-Schechter solutions.
The meaning of anomaly equation and applications to QCD and electroweak
theory are briefly discussed. We also comment on the relation of
the results of this paper to the previous work.
\vskip 1truecm
\vfil\eject

\baselineskip=6mm
\parindent 20pt
\centerline{\titlefnt 1. Introduction}

In the Standard Model fermion number is not conserved [1]
since gauge field
configurations with non-zero topological charge $Q$,
$$ Q = {g^2 \over 16 \pi^2}
\ \int d^4x~ {1\over 2}\epsilon^{\mu\nu\alpha\beta}{\rm tr}
  (F_{\mu \nu} F_{\alpha \beta}) \ ,
\eqno(1.1)$$
cause
violation of conservation laws due to anomalies [2].
In fact, the fermion number current-density
$\hat{j}^\mu = \hat{\overline\psi}_{\rm L}\gamma_\mu \hat\psi_{\rm L}$
for each left-handed fermion flavour is not conserved according to the
anomaly equation,
$$
\partial_\mu \hat{j}^\mu = {g^2 \over 16 \pi^2}
{1\over 2}\epsilon^{\mu\nu\alpha\beta}{\rm tr}
  (F_{\mu \nu} F_{\alpha \beta}) \ .
\eqno(1.2)$$

In a non-Abelian gauge theory there exist an unstable finite energy static
 solution
called the sphaleron [3]. It is a saddle-point of the gauge field potential
energy and its energy, $E_{\rm sp}$, is the barrier height between the
different
vacuum sectors of the theory.
When the transition between two different sectors occurs, the topological
charge
$Q$ of the gauge field interpolating between two different sectors is non-zero
and, according to the anomaly, eq. (1.2), the fermion number changes.
One way
of understanding these effects is to use semiclassical barrier penetration
approach where the tunneling solutions are Euclidean instantons [4].
In the electroweak theory the height of the barrier $E_{\rm sp}$  is  of order
$M_{\rm w}/\alpha_{\rm \scriptscriptstyle w} \sim 10 {\rm TeV}$
and at energies much below this, fermion number violation
is exponentially suppressed.
It has been suggested that fermion number violating processes
may become unsuppressed in the scattering processes at sufficiently
high energies [5]. An intuitive way to put it is that with increasing energy
the field should tunnel under smaller and smaller portion
of the barrier and at the energy higher than the barrier height
instead of tunneling through the
barrier the field configuration passes over it. Passage over the
sphaleron barrier is classically allowed and should be mediated
by a classical solution in Minkowski space-time. This is in contrast with the
tunneling process which is dominated by a classical solution in Euclidean
or even more generally complex time (for references and see a review [6]).

Minkowski space-time approach to fermion number violation may be
separated into three parts: the creation of finite energy
gauge field configurations by particle collisions, their classical evolution
with time and the dynamics of fermion
number (or chirality) violation in the presence of such classical
background gauge sources.
The classical evolution of certain gauge field
configurations in Minkowski space was addressed
in Refs. [7-8].
In this paper we study fermion number violation in the background
of a gauge field in Minkowski space. Then we apply our results to the case
of spherically symmetric classical gauge configurations in the background.

In the Euclidean approach the functional integral is dominated by
instantons or instanton-like configurations with finite action.
These configurations fall into homotopy classes and require the
topological charge $Q$ to be an {\it integer}.
Then according to the anomaly equation (1.2), the number of fermions
of each flavour is changed [1] by an integer amount $Q$.

On classical solutions in Minkowski space $Q$ in general can take any value,
not just an integer [7,8].
 This is a consequence of the fact that
classical Minkowski gauge fields do not approach just some pure gauges
in the far past and future, but the finite energy radiation is always present.
In this case when the topological charge of a classical
gauge field background
is not an integer one may ask what is the anomalous fermion production
in such a background. This is the motivation of the present paper.

The outline of the paper is as follows.
 In Section 2 we review the method of N. Christ [9] of studying
 fermion number violation
in a class of background gauge fields. Then we derive a formula
for a fermion number violation in such backgrounds and show that
it is always an integer.
 In Section 3 we first review results of Farhi, Khoze and Singleton [7]
 on classical
Yang-Mills system in the spherical ansatz.
 Then the  approach of the Section 2 is applied to the background gauge
 configurations
 which match classical solutions of Ref. [7] at all times $t$
 except the early past,
$t\to T_{\rm min}$, and the far future, $t\to T_{\rm max}$. At these times
we switch off the gauge invariant degrees of freedom of the background field.
This should correspond to the physical situation of interest where
an initial coherent gauge
field configuration was produced in the course of quantum collision
at some early
time, $T_{\rm min}$, and then evolved classically before decaying
into quantum radiation
at some late time, $T_{\rm max}$. The idea of our work is to calculate the
violation of the fermion number which occurred during the classical evolution
of the initial coherent state before it decayed. We assume here that there
were no fermion number violation before the coherent field was created or
after it decayed.
It will be seen  that the fermion number violation
which occurs in classical backgrounds is independent on the way of
how the interaction
is switched off at early and late times and neither it depends on the times
$T_{\rm min}$ and $T_{\rm max}$ as far as their absolute values
are much greater
than some characteristic time-scale, $|t_*|$,
 associated with the solutions. Thus, the fermion number
in our approach is indeed violated only during the classical evolution of the
initial coherent configuration and not at the moment of its creation or decay.
Moreover, we will see that it occurs at the instant where the background
field passes over some sphaleron-like configuration.
We will calculate the fermion number violation in the presence of classical
solutions explicitly and demonstrate that it is integer
and in general non-zero.

 The further questions which
arise from this work are discussed in Section 4.

\goodbreak
\bigskip

\centerline{\titlefnt 2. Fermion Number Violation }
\centerline{\titlefnt in the Background Gauge Field}

\bigskip

In this Section we present our interpretation of the approach of N.~Christ,
Ref. [9], Section IV C. Then we will demonstrate that the fermion number
is violated by at most an {\it integer} amount for a general class of
gauge field backgrounds.

For simplicity we consider the case of a single left-handed
fermion flavour
$\hat\psi_{\rm L}\equiv {1\over 2}\left( 1 - \gamma_5\right) \hat\psi$
coupled to an external $SU(2)$ gauge field.
The generalization for the fermion content
of the realistic theory is straightforward.
The Fermion operator $\hat\psi_{\rm L}$ obeys
$$
  i\gamma^\mu \left( \partial_\mu - i g A_\mu\right)
 \hat\psi_{\rm L} = 0 \ .
\eqno(2.1)
$$
 From now on we will suppress the L-subscript of the Fermi-fields bearing
in mind that all $\hat\psi$-s are left-handed. The hats distinguish
the operator-valued fields from the c-numbers.

We are interested here in background fields $A_\mu({\bf x},t)$
which in the early
past, $T_{\rm min}<t<T_{\rm i}\ll 0$, and in the far future,
$0 \ll T_{\rm f}<t<T_{\rm max}$, can be cast in the following form:
$$
   A_\mu ({\bf x},t<T_{\rm i}) =  U_{\rm in}({\bf x})
  \bigl[{i\over g} \partial_\mu + B_\mu^{\rm in}({\bf x},t)
  \bigr] U_{\rm in}^\dagger ({\bf x}) \ ,
\eqno(2.2a)
$$
$$
   A_\mu ({\bf x},t>T_{\rm f}) =  U_{\rm out}({\bf x})
  \bigl[{i\over g} \partial_\mu + B_\mu^{\rm out}({\bf x},t)
  \bigr] U_{\rm out}^\dagger ({\bf x}) \ .
\eqno(2.2b)
$$
Here $U_{\rm in}({\bf x})$ and $U_{\rm out}({\bf x})$ are $SU(2)$-valued
continuous functions of ${\bf x}$
which, as ${\bf x} \to \infty$,
approach direction-independent constants.
Thus, $U_{\rm in}({\bf x})$ and $U_{\rm out}({\bf x})$
can be characterized by  winding numbers, $\nu[U_{\rm in}]$
and $\nu[U_{\rm out}]$ which are integer numbers.
The gauge fields $B_\mu^{\rm in}({\bf x},t)$
and $B_\mu^{\rm out}({\bf x},t)$
on the right hand side of eqs. (2.2)
are required to have essentially {\it finite} support
in the ${\bf x}$-space at any fixed time
$ t$ and  to vanish at any ${\bf x}$ as time goes respectively to
$T_{\rm min}$ or $T_{\rm max}$.

We want to study a process of creation of fermions
in the background gauge field specified above.
Since we are concerned with particle creation it is important to be able
to distinguish a positive energy mode from a negative energy mode and
in general in order to count particles we would like to
have descrete energy levels.
For this reason we compactify the ${\bf x}$-space
at spatial infinity at any
{\it fixed} time $t$. This compactification is not in contradiction with
the gauge field backgrounds we consider, since $U_{\rm in}({\bf x})$
and $U_{\rm out}({\bf x})$
approach direction-independent constants at spatial infinity and
$B_\mu^{\rm in}({\bf x},t)$ and $B_\mu^{\rm out}({\bf x},t)$ are zero
at spatial infinity at a fixed time because of the essentially
finite support requirement.
In other words, when we'll have to deal with the order of
limits in Minkowski space-time,
our prescription will always be to first let the spatial
variable ${\bf x}$ go
to (compactified) infinity and then (if needed)
to let the time $T_{\rm min} < t < T_{\rm max}$
go to the infinite past, $T_{\rm min}\to -\infty$, or infinite future,
$T_{\rm max}\to +\infty$.

Our program now is to first find the Fermi-operator $\hat\psi({\bf x},t)$.
Then we can construct the fermion number operator
and consider its expectation values at $t=T_{\rm min}$ and $t=T_{\rm max}$.
The difference between these expectation values will give the
fermion number violation.

The Fermi-operator $\hat\psi({\bf x},t)$ is obtained by the procedure
of the second quantization from the  c-number general solution to
the equation of motion (2.1). To obtain this we have to find a complete
set of c-number solutions to (2.1).

We first consider a c-number solution of
$$
  i\gamma^\mu\left( \partial_\mu - i g A_\mu({\bf x},t) \right)
   \psi({\bf x},t) = 0 \ .
\eqno(2.3)$$
Let us make a gauge transformation with the gauge function
$U= U_{\rm in}({\bf x})$ of eq. (2.2a),
$$
   A_\mu ({\bf x},t) =  U_{\rm in}({\bf x})
  \bigl[{i\over g} \partial_\mu + B_\mu^{\rm in}({\bf x},t)
  \bigr] U_{\rm in}^\dagger ({\bf x}) \ ,
\eqno(2.4a) $$
$$
   \psi({\bf x},t) =  U_{\rm in}({\bf x}) \zeta({\bf x},t) \ .
\eqno(2.4b) $$
In terms of new variables $B_\mu^{\rm in}({\bf x},t)$ and
$\zeta({\bf x},t)$ eq. (2.3) reads
$$
  i\gamma^\mu\left( \partial_\mu - i g B_\mu^{\rm in}({\bf x},t) \right)
   \zeta({\bf x},t) = 0 \ .
\eqno(2.5)$$
Equations (2.4) describe the gauge in which
the background gauge field is $B_\mu^{\rm in}$ which
has to vanish at $t=T_{\rm min}$.
Thus, in this gauge fermions become free in the early past.

There is a complete orthonormal set of solutions of eq. (2.5)
which we call an in-set,
 $\bigl\{\zeta^{{\rm in}\pm}_n ({\bf x},t)\bigr\}_{n=1}^{\infty}$,
with the initial condition that
$$
  \zeta^{{\rm in}\pm}_n ({\bf x},t) \to \psi^\pm_n({\bf x})\, e^{\mp
  iE_nt} \ \ {\rm as} \ \ t\to T_{\rm min} \  \ .
\eqno(2.6)
$$
Here $\psi^\pm_n({\bf x})$ are positive and negative energy eigenfunctions
of the free Dirac Hamiltonian,
$$
  - i {\pmb \alpha}\cdot  {\pmb\nabla}\psi^\pm_n ({\bf x}) =
  \pm E_n \psi^\pm_n ({\bf x}) \ \ ,
\eqno(2.7)
$$
where ${\pmb \alpha}=\gamma^0{\pmb \gamma}$ and
$E_n>0$. (By a judicious choice of (compactified) boundary conditions for
fermion fields
at spatial infinity one can make each negative
energy equal to minus a positive energy and also eliminate the zero
energy eigenvalue.)

Equation (2.5) can be cast in the retarded Yang-Feldman form:
$$
 \zeta^{{\rm in}\pm}_n ({\bf x},t) = \psi^\pm_n({\bf x})\, e^{\mp iE_nt}
\ - \ g\int^t_{T_{\rm min}} dy_0 \int d{\bf y} \Delta^{\rm ret}(x-y)
 \ \gamma^\mu B_\mu^{\rm in}({\bf y},y_0) \ \zeta^{{\rm in}\pm}_n ({\bf y},y_0)
 \ ,  \eqno(2.8)$$
where the first term on the right hand side is the solution of the free Dirac
equation
and
$\Delta^{\rm ret}(x-y)$ is the retarded Green function,
$$
 i\gamma^\mu \partial_\mu \ \Delta^{\rm ret}(x-y) = \delta^{(4)} (x-y) \ ,
\eqno(2.9a)$$
$$
  \Delta^{\rm ret}(x-y) \sim \theta (x_0-y_0) \ .
\eqno(2.9b)$$
The (retarded) initial condition (2.6) is satisfied only for such
backgrounds $ B_\mu^{\rm in}$ that the integral on the right hand side
of eq. (2.8) vanishes as $t \to T_{\rm min}$.
If the equation (2.8) can be solved by iterations,
the in-set elements are given by the perturbative formula:
$$
 \zeta^{{\rm in}\pm}_n ({\bf x},t) = \psi^\pm_n({\bf x})\, e^{\mp iE_nt}
$$
$$
 - \ g\int^t_{T_{\rm min}} dy_0 \int d{\bf y} \Delta^{\rm ret}(x-y)
 \ \gamma^\mu B_\mu^{\rm in}({\bf y},y_0) \ \psi^\pm_n({\bf y})\,
 e^{\mp iE_n y_0}
\ + \ ...
 \ .  \eqno(2.10)$$
We can now finally return to our specification of $B_\mu^{\rm in}$ in the
beginning of the Section: $B_\mu^{\rm in}$ is required to vanish
as $t \to T_{\rm min}$ fast enough that the integral(s) on the right
hand side
of eq. (2.10) are well defined and vanish as $t \to T_{\rm min}$
and the solution
of eq. (2.8) by iterations makes sence\footnote{$^\clubsuit$}{This point
was investigated in Ref. [10]. What is rather important for our applications
in the next Section is the fact that the classical gauge field solutions
of Ref. [7] {\it cannot} be cast in the form
to allow iterations of the Yang-Feldman equation contrary to the claim
of Ref. [10]. We will return to this point in Section 3. Here we just note
that in order to apply the formalism of this Section to the case of
classical fields in the background, the background should be modified
at the early past and the far future to switch off the
interactions with fermions.}
In this case we also see that $\zeta^{{\rm in}+}_n ({\bf x},t)$ are positive
and $\zeta^{{\rm in}-}_n ({\bf x},t)$ are
negative frequency solutions as $t\to T_{\rm min}$ which will allow
a particle interpretation.

The general c-number solution to
the equation of motion (2.1) is an arbitrary linear
combination of the elements
of the complete in-set.
The Fermi-operator $\hat{\zeta}({\bf x},t)$ is obtained from this by
declaring the coefficients in front of the negative and positive
frequency components
to be the creation and
annihilation operators respectively,
$$
  \hat{\zeta} ({\bf x},t) = \sum_{n=1}^{\infty} \left[ \hat a^{\rm in}_n
  \zeta^{{\rm  in}+}_n
  ({\bf x},t) + \hat b^{{\rm in}\dagger}_n \zeta^{{\rm in}-}_n ({\bf
  x},t)\right] \ .
  \eqno(2.11)
$$
Here $\hat a^{\rm in}_n$ is the annihilation operator of a particle
with the energy $E_n$ in the in-state, while
$\hat b^{{\rm in}\dagger}_n$ is the creation operator of an anti-particle
with the energy $E_n$ in the in-state.
Since the integrals on the right hand side of equation (2.10) vanish
as $t \to T_{\rm min}$, these creation and annihilation
operators obey the usual (free) anti-commutation relations and
the in-vacuum state, $|0^{\rm in}\rangle$, is defined as:
$$
  \hat a^{\rm in}_n |0^{\rm in}\rangle = \hat b^{\rm in}_n |0^{\rm
  in}\rangle = 0 \ .
\eqno(2.12)
$$
Gauge transforming eq. (2.11) back to the original notations,
$$
  \hat{\psi} ({\bf x},t) = \sum_{n=1}^{\infty} \left[ \hat a^{\rm in}_n
  U_{\rm in}({\bf x}) \zeta^{{\rm  in}+}_n
  ({\bf x},t) + \hat b^{{\rm in}\dagger}_n U_{\rm in}({\bf x})
  \zeta^{{\rm in}-}_n ({\bf
  x},t)\right] \ ,
  \eqno(2.13)
$$
we obtain the Fermi-operator in the in-representation.

Our next goal is to obtain a representation of $\hat{\psi} ({\bf x},t)$
in terms of the {\it out-} creation and annihilation operators.
To do this we return to eq. (2.3) and
repeat the previous steps with certain modifications.
Consider  a gauge transformation
with the gauge function
$U= U_{\rm out}({\bf x})$ of eq. (2.2b),
$$
   A_\mu ({\bf x},t) =  U_{\rm out}({\bf x})
  \bigl[{i\over g} \partial_\mu + B_\mu^{\rm out}({\bf x},t)
  \bigr] U_{\rm out}^\dagger ({\bf x}) \ ,
\eqno(2.14a) $$
$$
   \psi({\bf x},t) =  U_{\rm out}({\bf x}) \xi({\bf x},t) \ .
\eqno(2.14b) $$
Equation (2.3) takes the form:
$$
  i\gamma^\mu\left( \partial_\mu - i g B_\mu^{\rm out}({\bf x},t) \right)
   \xi({\bf x},t) = 0 \ .
\eqno(2.15)$$
The background gauge field now is $B_\mu^{\rm out}$ which
has to vanish in the far future.
In this gauge fermions become free as $t \to T_{\rm max}$.

A complete orthonormal out-set of solutions of eq. (2.15),
 $\bigl\{\xi^{{\rm out}\pm}_n ({\bf x},t)\bigr\}_{n=1}^{\infty}$,
is defined by  the ``initial" condition,
$$
  \xi^{{\rm out}\pm}_n ({\bf x},t) \to \psi^\pm_n({\bf x})\, e^{\mp
  iE_nt} \ \ {\rm as} \ \ t\to T_{\rm max} \  \ .
\eqno(2.16)
$$
We now use the advanced Yang-Feldman form of the equation (2.15):
$$
 \xi^{{\rm out}\pm}_n ({\bf x},t) = \psi^\pm_n({\bf x})\, e^{\mp iE_nt}
\ - \ g\int_t^{ T_{\rm max}} dy_0 \int d{\bf y} \Delta^{\rm adv}(x-y)
 \ \gamma^\mu B_\mu^{\rm out}({\bf y},y_0) \
 \xi^{{\rm out}\pm}_n ({\bf y},y_0)
 \ ,  \eqno(2.17)$$
where
$\Delta^{\rm adv}(x-y)$ is the advanced Green function,
$$
 i\gamma^\mu \partial_\mu \ \Delta^{\rm adv}(x-y) = \delta^{(4)} (x-y) \ ,
\eqno(2.18a)$$
$$
  \Delta^{\rm adv}(x-y) \sim \theta (y_0-x_0) \ .
\eqno(2.18b)$$
 Now the (advanced) initial condition (2.16) is satisfied only for such
backgrounds $ B_\mu^{\rm out}$ that the integral on the right hand side
of eq. (2.17) vanishes as $t \to  T_{\rm max}$.
The out-set elements are given by the iterative solution of equation (2.17):
$$
 \xi^{{\rm out}\pm}_n ({\bf x},t) = \psi^\pm_n({\bf x})\, e^{\mp iE_nt}
$$
$$
 - \ g\int_t^{ T_{\rm max}} dy_0 \int d{\bf y} \Delta^{\rm adv}(x-y)
 \ \gamma^\mu B_\mu^{\rm out}({\bf y},y_0) \ \psi^\pm_n({\bf y})\,
 e^{\mp iE_n y_0}
\ + \ ...
 \ .  \eqno(2.19)$$
$B_\mu^{\rm out}$ is required to vanish as
$t \to  T_{\rm max}$ that the integral(s) on the right hand side
of eq. (2.19) are well defined and vanish{$^\clubsuit$}
 as $t \to  T_{\rm max}$.

The Fermi-operator $\hat{\xi}({\bf x},t)$ is
$$
  \hat{\xi} ({\bf x},t) = \sum_{n=1}^{\infty} \left[ \hat a^{\rm out}_n
  \xi^{{\rm  out}+}_n
  ({\bf x},t) + \hat b^{{\rm out}\dagger}_n \xi^{{\rm out}-}_n ({\bf
  x},t)\right] \ ,
  \eqno(2.20)
$$
with $\hat a^{\rm out}_n$ being the annihilation operator of a fermion
  and
$\hat b^{{\rm out}\dagger}_n$ being the creation operator of an anti-fermion
in the out-state.
Since the integrals on the right hand side of equation (2.20) vanish
as $t \to  T_{\rm max}$, these out- creation and annihilation
operators obey the usual (free) anti-commutation relations and
the out-vacuum state, $|0^{\rm out}\rangle$, is:
$$
  \hat a^{\rm out}_n |0^{\rm out}\rangle = \hat b^{\rm out}_n |0^{\rm
  out}\rangle = 0 \ .
\eqno(2.21)
$$
Gauge transforming eq. (2.20) back we obtain the Fermi-operator
in the in-representation,
$$
  \hat{\psi} ({\bf x},t) = \sum_{n=1}^{\infty} \left[ \hat a^{\rm out}_n
  U_{\rm out}({\bf x}) \xi^{{\rm  out}+}_n
  ({\bf x},t) +
  \hat b^{{\rm out}\dagger}_n U_{\rm out}({\bf x}) \xi^{{\rm out}-}_n ({\bf
  x},t)\right] \ .
  \eqno(2.22)
$$

Equations (2.13) and (2.22) give two different representations of
$\hat{\psi} ({\bf x},t)$
in terms of two complete sets,
$\bigl\{\zeta^{{\rm in}\pm}_n ({\bf x},t)\bigr\}_{n=1}^{\infty}$ and
$\bigl\{\xi^{{\rm out}\pm}_n ({\bf x},t)\bigr\}_{n=1}^{\infty}$,
given by equations (2.10) and (2.19).

We now construct the operator of the fermionic current-density,
$\hat{j}^\mu (x) = \hat{\overline\psi}(x) \gamma^\mu \hat\psi(x)$.
We remind that $\hat\psi(x)$ is the left-handed fermion, so
$\hat{j}^\mu (x)$ is a combination of an axial-vector and a vector
current-density. We will require the vector charge
to be conserved in the quantized theory
(the theory remains gauge invariant) and the axial-vector
charge will be violated anomalously.

The current-density operator, $\hat{j}^\mu (x)$, is a
composite operator built out of local operators at
the same space-time
point $x$. For the integrals of the current-density,
such as the charge operator, $\int d^3 x \hat{j}^0 (x)$,
to be regular, the composite operator $\hat{j}^\mu (x)$ should
be renormalized. The regularization should preserve gauge invariance.
We use the $\epsilon$-splitting regularization
of Schwinger and define the renormalized current-density as
$$
\hat{j}^\mu (x) = {\rm lim}_{\epsilon \to 0} \bigl(
\hat{j}^\mu (x|\epsilon) - \{ {\rm counter \ term} \}^\mu \bigr) \ ,
\eqno(2.23)$$
where the gauge invariant point-split current is
$$
\hat{j}^\mu (x|\epsilon)=
\hat{\overline\psi} (x+\epsilon/2)\gamma^\mu
\ {\cal P}\exp\bigl[ ig \int_{x-\epsilon/2}^{x+\epsilon/2}
dy^\nu A_\nu (y) \bigr] \
\hat\psi (x-\epsilon/2) \ ,
\eqno(2.24)$$
and the counter term is independent of the gauge field $A_\mu (x)$.
The $\int d^3 x \{ {\rm counter \ term} \}^0$ is a time-independent
infinite constant
to be subtracted from the unrenormalized charge operator to
make the charge of the vacuum finite. Since the counter term is
time-independent,
the effects of finite renormalization will cancel out in the difference
of the charges in the beginning and at the end of the day.

To give $\hat{j}^\mu (x|\epsilon)$ the correct properties under
Lorentz transformations, the limit $\epsilon \to 0$
should be taken symmetrically [11]:
$$
\epsilon^\mu \to 0 \ \ \ \
\epsilon^\mu \epsilon^\nu/\epsilon^2 \to g^{\mu \nu}/4 \ .
\eqno(2.25)$$
Symmetric limit means that we first average over directions
of $\epsilon$ and then let $\epsilon^2=\epsilon^\mu \epsilon_\mu \to 0$.

Using the in-representation of the Fermi-operator, eq. (2.13),
the $\epsilon$-split current-density we find
$$
 \hat{j}^\mu (x|\epsilon)=  \sum_{n=1}^{\infty} \left[
 \hat a^{{\rm in}\dagger}_n
  \overline{\zeta^{{\rm  in}+}}_n U_{\rm in}^{\dagger}
   + \hat b^{\rm in}_n \overline{\zeta^{{\rm in}-}}_n U_{\rm in}^{\dagger}
\right]|_{(x+\epsilon/2)} \cdot$$
$$
\gamma^\mu
\ {\cal P}\exp\bigl[ ig \int_{x-\epsilon/2}^{x+\epsilon/2}
dy^\nu A_\nu (y) \bigr] \
 \sum_{m=1}^{\infty} \left[ \hat a^{\rm in}_m
  U_{\rm in} \zeta^{{\rm  in}+}_m
   + \hat b^{{\rm in}\dagger}_m U_{\rm in} \zeta^{{\rm in}-}_m
\right]|_{(x-\epsilon/2)} $$
$$
  = :\hat{j}^\mu (x|\epsilon):_{\rm in} +
S_{\rm in}^{\epsilon \ \mu}[A] \ ,
\eqno(2.26)$$
where $:\hat{j}^\mu (x|\epsilon):_{\rm in}$ is the normal form
of $:\hat{j}^\mu (x|\epsilon):_{\rm in}$
with respect to the in- creation
and annihilation operators and
$$
S_{\rm in}^{\epsilon \ \mu}[A] =
\sum_{n=1}^{\infty}
\overline{\zeta^{{\rm in} -}_n} (x+\epsilon/2) \gamma^\mu
\ {\cal P}\exp\bigl[ ig \int_{x-\epsilon/2}^{x+\epsilon/2}
dy^\nu B^{\rm in}_\nu (y) \bigr] \
\zeta^{{\rm in}-}_n (x-\epsilon/2) \ .
\eqno(2.27)$$
Here we used
the anti-commutation relations and the gauge invariance
of the point-split construction.
The charge build from the normal ordered current-density
$:\hat{j}^\mu (x|\epsilon):_{\rm in}$
is regular in the $\epsilon \to 0$ limit.
Thus, the counter term can be chosen as follows:
$$ \{ {\rm counter \ term} \}^\mu =  S^{\epsilon \ \mu}[A\equiv 0]
\equiv
\sum_{n=1}^{\infty}
\overline{\psi^{-}_n} ({\bf x}+{\pmb \epsilon}/2)  \gamma^\mu
\psi^{-}_n ({\bf x}-{\pmb \epsilon}/2) \ ,
\eqno(2.28)$$
where $\psi^{-}_n ({\bf x})$ are negative energy eigenfunctions
of the free Dirac Hamiltonian, eq. (2.7). (As it should be, the counter
term is time-independent and does not depend on $A_\mu$.)

The operator,
$$\hat{N}_{\rm i} = {\rm lim}_{t \to
T_{\rm min}} \int d^3 x :\hat{j}^0 (x|0):_{\rm in}
 = \sum_{n=1}^{\infty} \bigl(
\hat a^{{\rm in}\dagger}_n \hat a^{\rm in}_n \ -
 \ \hat b^{{\rm in}\dagger}_n \hat b^{\rm in}_n \bigr) \ ,
\eqno(2.29)$$
measures the net fermion number in the early past.

Similarly,
the fermion number in the far future is given by
$$\hat{N}_{\rm f} = {\rm lim}_{t \to T_{\rm max}}
\int d^3 x :\hat{j}^0 (x|0):_{\rm out}
 = \sum_{n=1}^{\infty} \bigl(
\hat a^{{\rm out}\dagger}_n \hat a^{\rm out}_n \ -
 \ \hat b^{{\rm out}\dagger}_n \hat b^{\rm out}_n \bigr) \ ,
\eqno(2.30)$$
where $:\hat{j}^\mu (x|\epsilon):_{\rm out}$ is the normal
ordered current-density operator with respect to the out- creation
and annihilation operators and,
$$
\hat{j}^\mu (x|\epsilon) = :\hat{j}^\mu (x|\epsilon):_{\rm out} +
S_{\rm out}^{\epsilon \ \mu}[A] \ ,
\eqno(2.31)$$
where,
$$
S_{\rm out}^{\epsilon \ \mu}[A] =
\sum_{n=1}^{\infty}
\overline{\xi^{{\rm out}-}_n} (x+\epsilon/2) \ \gamma^\mu
\ {\cal P}\exp\bigl[ ig \int_{x-\epsilon/2}^{x+\epsilon/2}
dy^\nu B^{\rm out}_\nu (y) \bigr] \
\xi^{{\rm out} -}_n (x-\epsilon/2) \ .
\eqno(2.32)$$

The fermion number violation is the expectation value of
$$ \hat{N}_{\rm f} - \hat{N}_{\rm i} =
{\rm lim}_{t \to T_{\rm max}} \int d^3 x :\hat{j}^0 (x|0):_{\rm out} \ - \
{\rm lim}_{t \to T_{\rm min}} \int d^3 x :\hat{j}^0 (x|0):_{\rm in}
$$
$$
= \int_{T_{\rm min}}^{T_{\rm max}} dt \int d^3 x \
 {\rm lim}_{\epsilon \to 0} \partial_t
\hat{j}^0 (x|\epsilon) \
\eqno(2.33)
$$
$$
 - \ {\rm lim}_{\epsilon \to 0}\biggl[
{\rm lim}_{t \to T_{\rm max}} \int d^3 x \bigl( S_{\rm out}^{\epsilon \ 0}[A]
- S^{\epsilon \ 0} [0]\bigr) \ - \
{\rm lim}_{t \to T_{\rm min}} \int d^3 x \bigl( S_{\rm in}^{\epsilon \ 0}[A]
- S^{\epsilon \ 0} [0]\bigr) \biggr] \ .
$$
In deriving eq. (2.33) we used the fact that the counter term is
a time-independent constant.
The first term on the right hand side of eq. (2.33)
can be written as,
$$
\int_{T_{\rm min}}^{T_{\rm max}} dt \int d^3 x \
{\rm lim}_{\epsilon \to 0} \partial_t
\hat{j}^0 (x|\epsilon) =
\int d^4 x \ {\rm lim}_{\epsilon \to 0} \partial_\mu
\hat{j}^\mu (x|\epsilon) \ ,
\eqno(2.34)
$$
since the boundary terms at the surface at the spatial infinity
(at finite time, $T_{\rm min}<t<T_{\rm max}$) are vanishing.
As a result of a direct computation [11] we also have,
$$
 \int d^4 x \ {\rm lim}_{\epsilon \to 0} \partial_\mu
\hat{j}^\mu (x|\epsilon)
= {g^2 \over 16 \pi^2}
{1\over 2}\epsilon^{\mu\nu\alpha\beta}\int d^4 x {\rm tr}
  (F_{\mu \nu} F_{\alpha \beta})\equiv Q \ .
\eqno(2.35)$$
The expression above is obtained by differentiating the right
hand side of eq. (2.24),
making use of Dirac equation (2.3) and finally taking the symmetric limit
$\epsilon \to 0$ as prescribed by eq. (2.25). This way of obtaining
the expression on the right hand side of eq. (2.34)
can be viewed as a derivation of the anomaly equation (1.2).

The second term on the right hand side of eq. (2.33) can be abbreviated as
$-(q^{\rm out}-q^{\rm in})$. Here $q^{\rm out}$ and $q^{\rm in}$  are the
``fermion" charges of the radiating gauge fields $B_\mu^{\rm out}$
and $B_\mu^{\rm in}$ and have nothing to do with the actual number
of fermions.
They can be calculated by substituting iterative
solutions\footnote{$^\heartsuit$}{Equations
(2.17) and (2.18) should be iterated three times}
 of the Yang-Feldman equations
(2.17) and (2.8) into the expressions for $S$, eq. (2.32), (2.27)
and first
performing the integrations over the three-space in (2.33)
and only then letting
$t$ to go to the infinite future or infinite past.
The other order of limits would be inconsistent
with our set up (and would give zero result).
$q^{\rm out}$ and $q^{\rm in}$ were calculated by N. Christ [9],
$$
q^{\rm out}= {\rm lim}_{t \to +\infty} \int d^3 x K^0[B^{\rm out}] \ ,
\eqno(2.36a)$$
$$
q^{\rm in}= {\rm lim}_{t \to -\infty} \int d^3 x K^0[B^{\rm in}] \ ,
\eqno(2.36b)$$
where $K_0 [A]$ is a zeroth component of the topological current,
$$
K^\mu [A] ={g^2 \over 16 \pi^2} \int d^3 x
\epsilon^{ \mu \nu \alpha \beta} \ {\rm tr}\bigl(
A_\nu F_{\alpha \beta} -{2\over 3}
A_\nu A_\alpha A_\beta \bigr) \ ,
\eqno(2.37)$$
and
$$
\partial_{\mu}K^\mu =
{g^2 \over 16 \pi^2}
{1\over 2}\epsilon^{\mu\nu\alpha\beta}{\rm tr}
  (F_{\mu \nu} F_{\alpha \beta}) \ .
\eqno(2.38)$$

We note that $q^{\rm out}$ and $q^{\rm in}$ are gauge invariant
under small gauge transformations while large gauge transformations
would be inconsistent with our requirements on $B^{\rm out}$ and
$B^{\rm in}$ of falling off with time and should be absorbed into
$U_{\rm out}$ and $U_{\rm in}$.

Putting all the bits together, we reproduce N. Christ's
result [9]:
$$ \langle \hat{N}_{\rm f} - \hat{N}_{\rm i} \rangle =
Q - q^{\rm out} + q^{\rm in} \ .
\eqno(2.39) $$
Thus, when there is a radiation field, $B_\mu^{\rm (in)out}$, present
in the initial or final state, the net violation
of the classically conserved number of chiral fermions
is not given by the integral of the axial-vector anomaly
(topological charge $Q$),
but additional subtractions must be made [9].
The so-called fermionic
charge, $\int d^3 x \hat{j}^0$, contains a piece
$q^{\rm (in)out}$ which is the
``fermion" charge\footnote{$^\star$}{The $B_\mu^{\rm (in)out}$ fields
should go to zero as $t \to T_{\rm (min)max}$ in order to have
free fermions
at early and late times and iterate Yang-Feldman equations,
but this does not
guarantee that $q^{\rm (in)out}$ necessarily vanish due to the
order of limits
in eqs. (2.36)}
of the radiating gauge field $B_\mu^{\rm (in)out}$
 and has not much to do with the actual number of fermions which
in its turn is measured by a corresponding normal ordered product.

This is rather interesting since the topological charge $Q$
does not have to be an
integer [7] and
one may hope that the subtraction of $q^{\rm (in)out}$ will somehow
make the net effect of the fermion number violation to be
an integer\footnote{$^\sharp$}{It
would have been rather unpleasant to find a non-integer
number of fundamental fermions
in the detector at the end of a scattering experiment},

We will show now that
the Christ's result, eq. (2.39),
can be put in the form in which
the fermion number is always violated by an {\it integer} amount for
 arbitrary gauge field in the background
which allows iterations of the  Yang-Feldman equations (2.17), (2.8).
We have,
$$\langle \hat{N}_{\rm f} - \hat{N}_{\rm i} \rangle =
Q - q^{\rm out} + q^{\rm in}
$$
$$
=  {\rm lim}_{T \to +\infty} \int_{-T}^{T} dt \int d^3 x \
\partial_{\mu}K^\mu [A]
\ - \ {\rm lim}_{t \to +\infty} \int d^3 x K^0[B^{\rm out}] \ + \
{\rm lim}_{t \to -\infty} \int d^3 x K^0[B^{\rm in}] $$
$$
= \int d^3 x \  K^0\bigl[U_{\rm out}({\bf x}){i\over g}
\partial_\mu U_{\rm out}^\dagger ({\bf x})\bigr]
\ - \  \int d^3 x \  K^0 \bigl[U_{\rm in}({\bf x}){i\over g}
\partial_\mu U_{\rm in}^\dagger ({\bf x})\bigr]     $$
$$
 \equiv \nu [U_{\rm out}] - \nu [U_{\rm in}]
\ \in {\cal Z} \ ,
\eqno(2.40)$$
which is an integer since the winding numbers of $U_{\rm (in)out}$ are
integer by construction.

An important thing is to make sure that the integer baryon number violation
is not always zero for example on  Minkowski space classical solutions.
In the next Section we will calculate the fermion number violation
 in the background
of the spherical solutions [7].
We will demonstrate that it
is integer and non-zero in general and also derive some
useful selection rules.

\goodbreak
\bigskip

\centerline{\titlefnt 3. Classical Solutions in the Spherical Ansatz}
\centerline{\titlefnt and Fermion Number Violation}
\bigskip

Working in  the spherical ansatz
for pure $SU(2)$ gauge theory we will first review how [7]
the equations of motion
can be reduced to two equations for two gauge invariant variables
$\rho^2$ and $\psi$. Then we will discuss classical solutions in
(3+1)-dimensional
Minkowski space and calculate the violation of the fermion
number in their background.

The action for pure $SU(2)$ gauge theory is
$$
  S =  -{1\over 2} \int d^4 x \ {\rm tr} \left( F_{\mu\nu}F^{\mu\nu}\right)
,\eqno(3.1)$$
where $F_{\mu\nu}= F^a_{\mu\nu} \left(\sigma^a/2\right) = \partial_\mu
A_\nu - \partial_\nu A_\mu - i g \left[ A_\mu, A_\nu\right]$ is the field
strength and $A_\mu= A^a_\mu\left(\sigma^a/2\right)$.

The spherical  ansatz [12] is given in terms of the four functions
$a_0,a_1,\alpha,\beta$ by
$$\eqalign{
  A_0 ({\bf x},t) &= {1\over 2g} ~ a_0 (r,t) \ {\pmb\sigma}\cdot
  \hat{\bf x}\ \ ,\cr
  A_i({\bf x},t) &= {1\over 2g}\  \left( a_1 (r,t) \
  {\pmb\sigma} \cdot\hat{\bf x} \hat x_i +
  {\alpha(r,t)\over
  r} \ (\sigma_i - {\pmb\sigma} \cdot \hat{\bf x}\hat x_i)  +
  {1+\beta(r,t)\over r} \epsilon_{ijk} \hat x_j \sigma_k \right)\ \ ,\cr}
  \eqno(3.2)$$
 where $\hat{\bf x}$ is a unit three-vector in the radial direction.
Note that $1/g$ factors are introduced in eqs. (3.2) as was done in
Refs. [12,7] which makes equations of motion  $g$-independent. This was not
so in the treatment of Ref. [8]. Perturbative solutions of Ref. [8] will
be mentioned in the next Section.

The action (3.1) in the spherical  ansatz takes the form
$$  S = {4\pi\over g^2} \int dt \int^\infty_0 dr \left( -{1\over 4} r^2
  f_{\mu\nu} f^{\mu\nu} - \left( D_\mu \chi\right)^* D^\mu \chi - {1\over
  2r^2}\left(|\chi|^2 - 1 \right)^2\right)\ \ .\eqno(3.3)$$
where $f_{\mu\nu} =
\partial_\mu a_\nu - \partial_\nu a_\mu$ with $\mu,\nu=t,r$,
is the $(1+1)$-dimensional field strength,
$\chi = \alpha+i\beta$ is a complex scalar and
$D_\mu
\chi = \left( \partial_\mu - i a_\mu\right)\chi$ is the covariant derivative.
 To keep up with notations of Ref. [7], in the spherical ansatz
indices are raised and lowered with the $1+1$ dimensional metric
$\eta_{\mu\nu}={\rm diag}(-1,+1)$.

The  ansatz (3.2) preserves a residual $U(1)$ subgroup of the $SU(2)$
 gauge group consisting of the transformations,
$$  U({\bf x},t) = \exp \left[i\Omega(r,t) {{\pmb \sigma}\cdot\hat{\bf
  x}\over 2} \right]\  .\eqno(3.4)$$
These induce the gauge transformations
$$  a_\mu \to a_\mu + \partial_\mu \Omega\ \ ,\qquad \chi \to
\exp (i\Omega)
  \chi \ \ ,\eqno(3.5)$$
which leave (3.3) invariant.

The $(1+1)$-dimensional equations of motion for the reduced theory (3.4)
are given by
$$  - \partial^\mu \left(r^2 f_{\mu\nu} \right) = i \left[ \left( D_\nu
  \chi\right)^*\chi - \chi^* D_\nu \chi\right]\ ,\eqno(3.6\hbox{a})$$
$$  \left( - D^2 + {1\over r^2} \left( |\chi|^2-1\right)\right) \chi = 0
\ . \eqno(3.6\hbox{b})$$
Let us express the complex scalar field $\chi$ in polar form,
$$  \chi(r,t) = -i \rho(r,t) \exp \left[ i \varphi(r,t)\right]\ \
,\eqno(3.7)$$
where $\rho$ and $\varphi$ are real scalar fields and $\rho(r,t) \ge 0$.

One must bear in mind that in a point $\rho$ where vanishes the
angle $\varphi$ is not defined.
Assume that $\rho$ vanishes at a single point ($r_{*}$,$t_{*}$).
Surround the point ($r_{*}$,$t_{*}$) by a simple closed contour
in the $(r,t)$-space.
Then, since  $\chi$ is continuous and
$\rho \neq 0$ on the contour,
the change of $\varphi$ along the contour is in general an integer multiple
of $2 \pi$.
This integer multiple will be called a {\it degree} of $\varphi$ in the point
($r_{*}$,$t_{*}$).
Degree of $\varphi$ is non-zero only if $\varphi$ changes discontinuously
in the point ($r_{*}$,$t_{*}$) which is called then a singular point.

One of the central results of this Section will be a derivation of
the selection rule:
{\it the change of the numbers of fermions is equal to the sum of
the degrees of
$\varphi$ in each singular point}.
This is an integer by construction (which cannot [7] be said about
the topological charge).

In terms of $\rho$, $\varphi$ and $a_\mu$, the four equations contained
in (3.6) read
$$
  \partial^\mu\left( r^2 f_{\mu\nu}\right) + 2\rho^2 \left(
  \partial_\nu\varphi  - a_\nu\right) = 0\ \ , \eqno(3.8\hbox{a})
$$
$$
  \partial^\mu \partial_\mu \rho - \rho\left( \partial^\mu \varphi-
  a^\mu  \right)\left( \partial_\mu \varphi -a_\mu  \right)
 - {1\over r^2} \rho \left(\rho^2-1\right) = 0 \ \ ,
\eqno(3.8\hbox{b})
$$
and
$$
  \partial^\mu \big[\rho^2(\partial_\mu \varphi - a_\mu)\big]=0 \ .
\eqno(3.8\hbox{c})
$$
The last equation follows from (3.8a) so  there are three, not four,
independent equations, as expected because of the residual
$U(1)$ gauge invariance.

In practice the new field $\rho \equiv \sqrt{\alpha^2 +\beta^2}$
is not very convenient since it involves the square root of
the old variables.
It will be more useful for us to use $\rho^2 = \alpha^2 +\beta^2$
as the new primary field variable instead of $\rho$.
By rewriting eq. (3.8b) as
$$
 {1 \over 2} \partial^\mu \partial_\mu \rho^2
 - {1 \over 4 \rho^2} (\partial^\mu \rho^2) (\partial_\mu \rho^2)
 - \rho^2\left( \partial^\mu \varphi-
  a^\mu  \right)\left( \partial_\mu \varphi -a_\mu  \right)
 - {\rho^2 \over r^2}  \left(\rho^2-1\right) = 0 \ \ ,
\eqno(3.8\hbox{b'})
$$
we ensure that only $\rho^2$ and not $\rho$ appears in the
classical equations.

Since in (1+1) dimensions $f_{\mu\nu}$ must be proportional to
$\epsilon_{\mu\nu}$, we define [7] a new field $\psi$ as follows:
$$  r^2 f_{\mu\nu} = - 2 \epsilon_{\mu\nu}\psi\ \ ,\eqno(3.9)$$
here $\epsilon_{01}=+1$.
Equation (3.8a) now becomes
$$  \partial^\alpha \psi = - \epsilon^{\alpha\nu}\rho^2(\partial_\nu
  \varphi - a_\nu)\ \ .\eqno(3.10)$$
which implies
$$  \partial_\alpha\bigg({\partial^\alpha \psi \over \rho^2}\bigg) -
  {2\over r^2} \psi = 0 \ .
\eqno(3.11)$$
This gives an
equation solely in terms of the fields $\rho^2$ and $\psi$.
We may also use
(3.10) to express the second term in (3.8b') in terms of only $\rho^2$
and $\psi$,
$$  -\partial_t^2\rho^2 +\partial_r^2\rho^2
  + {1\over 2\rho^2}~\bigl( (\partial_t\rho^2)^2
  - (\partial_r\rho^2)^2\bigr)
 - {2\over \rho^2}~\bigl((\partial_t\psi)^2
  - (\partial_r\psi)^2 \bigr)
  - {2 \rho^2 \over r^2}(\rho^2-1) = 0 \ ,
\eqno(3.12\hbox{a})
$$
$$
  -\partial_t \big({\partial_t \psi \over \rho^2} \big)
  +\partial_r \big({\partial_r \psi \over \rho^2} \big)
  - {2\psi \over r^2} = 0 \ .
\eqno(3.12\hbox{b})
$$
Equations (3.12)  are equivalent to the original eqs.
(3.6), but now the fields are
$\rho^2$ and $\psi$ which are gauge invariant, and there are only two
equations in (3.12).

Using the equations of motion, the energy associated with the action
(3.3) can be written in terms of $\rho^2$ and $\psi$ as
$$
  \eqalign{E &= {8\pi \over g^2}~
  \int^\infty_0 dr\biggl[ {1\over 8\rho^2} \left(\partial_t
  \rho^2\right)^2 + {1\over 8\rho^2}\left( \partial_r \rho^2\right)^2
  +{1\over 2\rho^2} \left( \partial_t\psi\right)^2 \cr &\qquad+ {1\over
  2\rho^2} \left(\partial_r\psi\right)^2 + {\psi^2\over r^2} + {\left(
  \rho^2-1\right)^2\over4r^2}~\biggr]\ \ .\cr}
\eqno(3.13)$$
We are interested in finite energy solutions to
(3.12).

Witten [12] observed that (3.3) is the action for an Abelian Higgs
model in a curved space-time.
In fact [7],  the space-time manifold is the two dimensional
De Sitter space,
i.e.  hyperboloid $z_0^2-z_1^2-z_2^2=-1$ where the $z_i$
are functions of $r$ and $t$ and the coordinates $r$ and $t$ cover only
half of the hyperboloid for which $z_0+z_2 > 0$.
It is rather convenient to work with coordinates
$w$ and $\tau$ that live on the  hyperboloid.
The coordinate $w$ is a
bounded measure of the vertical position along the hyperboloid,
 $|w|< \pi/2$, and $\tau$ measures the azimuthal angle,  $|\tau| \le \pi$.
For more details see Fig. 1 of Ref. [7].
The explicit representation of $w$ and $\tau$ is given by
$$\eqalignno{
 & w = \ {\rm arctan}\bigl({1+t^2-r^2 \over 2 r}\bigr) \ ,
 &(3.14\hbox{a})  \cr
 & \tau = \ {\rm sign}(\tau) \
{\rm arccos}\bigl({1-t^2+r^2 \over \sqrt{(1+t^2-r^2)^2+4r^2}}\bigr) \ .
&(3.14\hbox{b})  \cr
  }
$$

 In terms of $w$-$\tau$ variables equations of motion (3.12) take
 the form [7]
$$
  -\partial_\tau^2\rho + \partial_w^2\rho
+ {\left(\partial_\tau \rho^2 \right)^2 \over 2\rho^2}
- {\left(\partial_w \rho^2 \right)^2 \over 2\rho^2}
- {2\left(\partial_\tau \psi \right)^2 \over \rho^2}
+ {2\left(\partial_w \psi\right)^2 \over \rho^2}
- {2\rho^2 \left(\rho^2-1\right) \over \cos^2 w} = 0 \ ,
\eqno(3.15\hbox{a})
$$
$$
  -\partial_\tau \left({\partial_\tau \psi \over \rho^2}\right) +
  \partial_w  \left({\partial_w\psi \over \rho^2}\right) -
  {2\psi \over \cos^2 w} = 0 \ .
\eqno(3.15\hbox{b})
$$

As a characteristic example of finite energy solutions to equations
of motion (3.15)
we consider solutions of L\"uscher and Schechter [13].
These solutions have finite energy, finite action and non-trivial
topological charge [7].
As was shown in Ref. [8], these explicit solutions are examples
of a wide class
of finite energy solutions all of which have certain general
features in common.
At early times they
depict a thin spherical shell of energy imploding towards the origin at near
the speed of light.  At around zero time the region around the origin is
energetically excited and at late times the shell is expanding outward,
asymptotically approaching the speed of light.

The main advantage of L\"uscher - Schechter solutions is that they
are known analytically:
$$\eqalignno{
  \rho^2(w,\tau) &= 1 + q(\tau) \ \bigl(q(\tau)+2\bigr)\cos^2 w \
  ,&(3.16\hbox{a})\cr
  \psi(w,\tau)   &= {1\over 2} ~ {dq(\tau)\over d\tau} \ \cos^2 w \
  , &(3.16\hbox{b})}$$
where the function
$q(\tau)$ is a solution of the ordinary differential equation:
$$\ddot q + 2q(q+1)(q+2)=0 \ .\eqno(3.17)$$
The  mechanical problem associated with eq. (3.17) is that of
a classical particle
trapped in the double well potential
$U={1\over 2} q^2\hbox{(q+2)}^2$.
The ``energy'' $\varepsilon$ of the ``particle" is
$$   \varepsilon = {1\over 2} \dot q^2 + U(q)\ . \eqno(3.18)$$
General solution of (3.17) will depend on the ``energy'' $\varepsilon$
and the ``time''-translation parameter $\tau_0$.
There are two classes of solutions depending on whether $\varepsilon$
is smaller or larger than $1/2$, the barrier height of $U(q)$ at
the unstable point $q=-1$:
$$\eqalign{
  q(\tau) &= -1 \pm (1+\sqrt{2\varepsilon})^{1/2} {\rm dn}\big(
  (1+\sqrt{2\varepsilon})^{1/2}(\tau-\tau_0) \ | \ m_1\big) \cr
  m_1 &=2\sqrt{2 \varepsilon}/(1+\sqrt{2\varepsilon});\qquad \varepsilon
  \le 1/2\ \ ,}\eqno(3.19)$$
and
$$\eqalign{
  q(\tau) &= -1 \pm (1+\sqrt{2\varepsilon})^{1/2} {\rm cn}\big(
  (8\varepsilon)^{1/4}(\tau-\tau_0) \ | \ m_2\big) \cr
 m_2 &=(1+\sqrt{2\varepsilon})/(2\sqrt{2 \varepsilon});\qquad
  \varepsilon
  > 1/2\ \ ,}\eqno(3.20)$$
where ${\rm dn}(u|m)$ and ${\rm cn}(u|m)$ are the Jacobi elliptic
functions\footnote{$^\ddagger$}{Since
there are several incompatible conventions in common mathematical use we will
always be using here notations of {\it Mathematica} [14]}
$$\eqalignno{
  u &= \int^1_{{\rm dn}(u|m)} {dt \over \sqrt{(1-t^2)(t^2+m-1)}}  \
  ,&(3.21\hbox{a})\cr
  u &= \int^1_{{\rm cn}(u|m)} {dt \over \sqrt{(1-t^2)(mt^2-m+1)}}
 \ , &(3.21\hbox{b})}$$
 There are always two forms of solutions ($\pm$ signs in (3.18) and (3.19))
since eq. (3.17) is not changed by the substitution $q=1+\kappa \to 1-\kappa$.
In particular when $\varepsilon<1/2$, different signs in eq. (3.19)
correspond to the particle being trapped in different wells.
The parameter $\tau_0$ corresponds to the
time at which the particle moving in the potential $U(q)$ with
energy $\varepsilon$ is at a turning point.

The L\"uscher-Schechter solutions can also be represented
in terms of the four original functions of the spherical ansatz
$$\eqalign{
  a_\mu &= -q(\tau)~\partial_\mu w  \ ,\cr
  \alpha &= {1\over 2} q(\tau)~ \sin 2w   \ , \cr
  \beta &=-(1 + q(\tau)\cos^2 w)\ , \cr}\eqno(3.22)$$
where $\mu=t,r$.

The L\"uscher-Schechter solutions
give spherically symmetric waves of localized energy density.
Now we would like to discuss the solution itself, i.e. $\rho^2 (r,t)$
and $\psi(r,t)$.
Figures 1 and 2 show the $r$-profiles of  $\rho^2(r,t)$ and $\psi(r,t)$
given by eqs. (3.16) for
a sequence of negative and positive times for a specific case of
 $\tau_0=1$ and $\varepsilon=1$.
In the distant past the ``two-dimensional" fields $\rho^2 (r,t)$ and
$\psi(r,t)$
are the incoming wave packets in the $r$-space which propagate
undistorted in a soliton-like manner at near the
speed of light. At around zero time the packets distort, collapse
and bounce back
producing outgoing wave packets. At large enough positive time the
outgoing wave packets
again propagate undistorted approaching the speed of light.
These $\rho^2 (r,t)$- and $\psi(r,t)$- packets represent imploding or expanding
spherical shells in (3+1) dimensions.
As the shell expands it leaves the
region of space behind it in a pure gauge configuration.
In the (1+1) dimensional $(r,t)$-space the outgoing wave packets
move undistorted.

As we already pointed out, these are the properties of not just
L\"uscher - Schechter solutions, but of a wide class of spherically symmetric
solutions [8].
Indeed, consider equations (3.12) and imagine that at some early time
$t=T_{\rm i} \ll 0$ the fields $\delta \equiv \rho^2-1 $ and $\psi$
are both pulses of width $\Delta$ centered at $r$ near $|T_{\rm i}|$ with
$\Delta \ll |T_{\rm i}|$. By a pulse we mean here a function which
is very close to zero
except in a region of the size $\Delta$.
For $r\sim  |T_{\rm i}| \gg \Delta$ we can now neglect the $1/r^2$
terms in eqs. (3.12).
We then see that if $\psi (r,t)$ and $\delta (r,t) \equiv \rho^2 (r,t) -1 $
depend
only on $r+t$, that is $\psi(r,t) = \psi_{\rm p}(r+t)$ and
$\delta(r,t) \equiv \rho^2 (r,t) -1 = \delta_{\rm p}(r+t)$ then eqs. (3.12)
are satisfied. Since $ \psi_{\rm p}(u)$ and $ \delta_{\rm p}(u)$
are close to zero
except for $u\sim \Delta$, the solution $\psi(r,t)$ and
$\rho^2(r,t)$ describe
incoming wave packets of the width $\Delta$ moving undistorted
along $r=-t$.
This description remains valid for all $t\ll -\Delta$.

At the late time $t=T_{\rm f} \gg 0$ the $1/r^2$ terms in
eqs. (3.12) can be
neglected again and the solution is described by pulses again,
$\psi(r,t) = \tilde\psi_{\rm p}(r-t)$ and
$\delta(r,t) \equiv  \rho^2 (r,t)-1 = \tilde\delta_{\rm p}(r-t)$
where $\tilde\psi_{\rm p}(v)$ and $\tilde\delta_{\rm p}(v)$ are
some new pulses of a width $\Delta$
and this
is valid for all $t\gg \Delta$.

We now return to Fig. 1 since there is one more important
lesson to be learned from L\"uscher - Schechter solutions.
It is apparent from Fig. 1a that there is a point in the $(r,t)$-space,
$(r_*,t_*)$, such that $\rho^2(r_*,t_*) =0$. We will show now that
$\varphi$ does change discontinuously
in the point ($r_{*}$,$t_{*}$) and the {\it degree} of $\varphi$
in the point
($r_{*}$,$t_{*}$) is 1.

It follows from equation (3.16a) that the $\rho^2$-component
of an arbitrary L\"uscher - Schechter solution can vanish at
the point $\tau_* = \tau (r_*,t_*)$, $w_* = w (r_*,t_*)$
in the $(\tau, w)$-space if and only if:
$$ q(\tau_*)=-1 \ , \eqno(3.23a)$$
$$ \cos^2 w_* = 1 \ . \eqno(3.23b)$$
The first condition can be satisfied only if $\varepsilon \ge 1/2$
since $q=-1$ is the height of the barrier, $U(q=-1)=1/2$.
Thus, solutions of the class (3.19) have non-vanishing $\rho^2$
and a continuous $\varphi$ at any $r$ and $t$. These solutions
describing a ``particle" trapped in a well will not cause
fermion number violation.
We now turn to solutions of the class (3.20).
The condition (3.23a) implies,
$$ \tau_{* n} = \tau_0 + {1 \over (8 \varepsilon)^{1/4} } u_{* n} \ ,
\eqno(3.24)$$
where $u= u_{* n}$ with $n=-\infty..\infty$ are the roots of
$ {\rm cn}( u_{* n} | m_2)$ given by
$$  u_{* n} = \int_0^{\pi/2 +\pi n} {d \theta \over
\sqrt{ 1-m_2 \sin^2 \theta}} =
(2n+1) \ {\rm K}(m_2) \ .
\eqno(3.25)$$
Solving conditions (3.23b) and (3.25) with the help of eqs. (3.14),
we have:
$$
r_{* n} = \sqrt{ 1+ t_{*n}^2} \ , \eqno(3.26a)$$
$$
t_{*n} = \tan \biggl(\tau_0 +
{1+2n \over (8 \varepsilon)^{1/4} } \  {\rm K} \bigl(
{1+\sqrt{2\varepsilon}\over 2\sqrt{2 \varepsilon}} \bigr) \biggr) \ ,
\eqno(3.26b)$$
$$ n \ : \quad -{\pi \over 2} \le
\tau_0 +
{1+2n \over (8 \varepsilon)^{1/4} } \  {\rm K} \bigl(
{1+\sqrt{2\varepsilon}\over 2\sqrt{2 \varepsilon}} \bigr)
 \le {\pi \over 2} \ .
\eqno(3.26c)$$
Equation (3.26c)
ensures that there is a certain {\it finite} number of times
$\rho^2$ vanishes, in particular, for the case of $\tau_0=1$
and $\varepsilon=1$, there is only one $n$ allowed by eq. (3.26c),
which is $n=-1$. This gives a single point $(r_{-1 *},t_{-1 *}) \simeq
(1.099, -0.455)$, which is consistent with Fig. 1a.

In general, $\rho^2$ vanishes each time the ``particle" of the
mechanical system
(3.17)-(3.18) goes over the top of the potential at $q=-1$ which we can call
the ''spahaleron of the double well".
Since the ``time" coordinate, $\tau$, of the mechanical analog is not
the time $t$ of the real world, but has a compact support on the
hyperboloid (3.14), the ``particle" goes through the ``sphaleron"
only a {\it finite}
number of times (determined by eq. (3.26c)), each time
approaching it from the different side.
We will see that each time this happens, the fermion number is violated by
$\pm 1$.

In fact with some algebra one can see that for an arbitrary
L\"uscher - Schechter solution
$\varphi$ changes discontinuously in each $(r_{* n},t_{* n})$
of eqs. (3.26)
and the {\it degree} of $\varphi$ in each of these points is $\pm 1$.
This can be proven by expanding $\alpha$ and $\beta$ around the
$(r_{* },t_{* })$-point,
$$
\alpha \sim -(r-r_*) +{t_* \over r_*} (t-t_*) \ , \eqno(3.27a)$$
$$
\beta \sim \dot q(\tau_*) {t_* \over r_*} (r-r_*) -
\dot q(\tau_*) (t-t_*) \ ,
\eqno(3.27b)$$
and evaluating the winding of the polar angle of $\alpha + i \beta$
along an infinitesimal circle around the $(r_*,t_*)$-point.

We are now ready to consider the fermion number violation in the presence
of the classical solutions in the spherical ansatz.
We assume that the classical equations (3.12) are solved and the fields
$\rho^2(r,t)$ and $\psi(r,t)$ are known.
In order to obtain the (3+1) dimensional form of the solution,
$A_\mu({\bf x},t)$
we have to find $a_0 (r,t)$, $a_1 (r,t)$, $\alpha(r,t)$ and $\beta(r,t)$
in terms of $\rho^2(r,t)$ and $\psi(r,t)$ in a given gauge.
 From eq. (3.10) we have
$$ \partial_t \psi  =
(a_1 - \partial_r \varphi) \ \rho^2 \ ,
\eqno(3.28a)
$$
$$ \partial_r \psi  =
(a_0 - \partial_t \varphi) \ \rho^2 \ .
\eqno(3.28b)
$$
If $\rho^2$ was non-vanishing at any $r$ and $t$ we could make
$\varphi(r,t)=0$ at any $r$ and $t$ by a continuous gauge transformation.
In this case we would have $a_0= \partial_r \psi / \rho^2$ and
$a_1= \partial_t \psi / \rho^2$ at any $r$ and $t$.
Such continuous gauge transformation $\Omega(r,t)=-\varphi(r,t)$
does not exist if $\varphi$ changes discontinuously when $\rho^2$ goes through
zero.

Suppose now that there is a single singular point $(r_*,t_*)$ where
$\rho^2$ vanishes and $\varphi$ changes discontinuously.
For definitness we start with a gauge $a_0=0$.
In this gauge we have from eq. (3.28b):
$$ \varphi(r,t)= -\int_{{\cal C}_{(r,T_{\rm min})\mapsto (r,t)} }
{\partial_r \psi \over \rho^2} \ dt \ ,
\eqno(3.29)$$
where we also put $ \varphi(r,T_{\rm min})=0$ by exhausting
the initial gauge freedom.
The contour of integration ${\cal C}_{(r,T_{\rm min})\mapsto (r,t)}$ runs
from $T_{\rm min}$ to $t$ surrounding the
singularity $(r_*,t_*)$ on the left as shown on Fig. 3.
The polar angle variable $ \varphi(r,t)$ of eqn (3.29)
is discontinuous on a ray $\{t=t_*, r \geq r_* \}$.
We can still make a continuous gauge transformation
(3.4) with some $\Omega_1 (r,t)$
which will make  $\varphi(r,t)=0$ at $t\ll t_*$ and all $r$.
For example we can chose
$$ \Omega_1 (r,t)= \int_{T_{\rm min}}^t
{\partial_r \psi(r,\tau) \over \rho^2(r,\tau) +h(\tau)} d\tau \ ,
\eqno(3.30a)$$
with $h(\tau)$ being a positive function with a support only at $\tau\sim t_*$.
Thus, $h$ makes $ \Omega_1$ well defined at $(r_*,t_*)$ and continuous,
but can be dropped at all $t\ll t_*$. We will call this
(specified by $ \Omega_1$) gauge
an initial gauge.

On the other hand the gauge $a_0=0$, $ \varphi(r,T_{\rm min})=0$ can be related
by a different continuous gauge transformation
$ \Omega_2 (r,t)$ with what will be called
a final gauge in which $\varphi(r,t)=0$ at $t\gg t_*$ and all $r$.
We chose
$$ \Omega_2 (r,t)= -\int^{T_{\rm max}}_t
{\partial_r \psi(r,\tau) \over \rho^2(r,\tau) +h(\tau)} d\tau
+ \int_{{\cal C}_{(r,T_{\rm min})\mapsto (r,T_{\rm max})} }
{\partial_r \psi(r,\tau) \over \rho^2(r,\tau)} \ d\tau
 \ .
\eqno(3.30b)$$

Finally, we relate the initial gauge in which $\varphi(r,t)=0$ at $t\ll t_*$
with the final gauge in which $\varphi(r,t)=0$ at $t\gg t_*$
by the gauge transformation
$ \Omega_{\rm f} (r)=
\Omega_2 (r,t)-\Omega_1 (r,t)$.

In the initial gauge
 the vector potential at early and late times is given by:
$$
   A_\mu ({\bf x},t\ll t_*) = B_\mu({\bf x},t) \ ,
\eqno(3.31a)
$$
$$
   A_\mu ({\bf x},t\gg t_*) =  U_{\rm f}({\bf x})
  \bigl[{i\over g} \partial_\mu + B_\mu({\bf x})
  \bigr] U_{\rm f}^\dagger ({\bf x},t) \ .
\eqno(3.31b)
$$
Here the field $B_\mu({\bf x},t)$ is just the right hand side
of eqs. (3.2) with
$$
a_0 (r,t) = \partial_r \psi(r,t) / \rho^2 (r,t) \ ,
\eqno(3.32a)$$
$$a_1(r,t) = \partial_t \psi(r,t) / \rho^2(r,t) \ ,
\eqno(3.32b)$$
$$ \alpha(r,t) = 0 \ ,
\eqno(3.32c)$$
$$ 1+\beta(r,t) = 1 - \rho(r,t) \ .
\eqno(3.32d)$$
The  gauge transformation $U_{\rm f}({\bf x},t)$ is an  $SU(2)$-valued
continuous function of ${\bf x}$,
$$U_{\rm f}({\bf x})=
\exp \left[i \Omega_{\rm f} (r) {{\pmb \sigma}\cdot\hat{\bf
  x}\over 2} \right]\  ,\eqno(3.33)$$
where
$$ \Omega_{\rm f} (r)=
-\int_{T_{\rm min}}^{T_{\rm max}}
{\partial_r \psi(r,\tau) \over \rho^2(r,\tau) +h(\tau)} d\tau
+\int_{{\cal C}_{(r,T_{\rm min})
\mapsto (r,T_{\rm max})}}
{\partial_r \psi \over \rho^2} d\tau \ .
\eqno(3.34)$$
First, we notice that $U_{\rm f}({\bf x})$ is, in fact, $t$-independent
since the contours of integration on the right hand side of eq. (3.34)
are $t$-independent.
We also note that, since the $\psi$ wave packets are
localized in the vicinity of the light-cone,
$ \Omega_{\rm f} (r)= 2\pi \cdot degree (\varphi(r_{*},t_{*}))$
and
$U_{\rm f}({\bf x}) = 1 $ for $ |{\bf x}| \gg {\rm max} (T_{\rm max},
|T_{\rm min}|)$.
Thus, $U_{\rm f}({\bf x})$ defines a mapping of a three-sphere into a
three-sphere which can be characterized by an integer winding number
$\nu(U_{\rm f})$ which is equal to  ${\it degree}(\varphi(r_{*},t_{*}))$.

A practical example for the discussion above is
a special case of a L\"uscher - Schechter solution with $\tau_0=1$
and $\varepsilon=1$, which has only one singular point
$(r_{-1 *},t_{-1 *}) \simeq (1.099, -0.455)$, given by eqs. (3.26).
Degree of  $ \varphi(r,t)$ in this point is 1 (and it can also be checked
explicitly that the right hand side of eq. (3.29) changes discontinuously
by $2 \pi$ in this point).

Now we have to consider the violation of the fermion number
in the presence of a background gauge field of eqs. (3.31)-(3.34).
Using the formalism of Section 2, fermion number violation can be calculated
in the presence of the background of the type (2.2).
The ansatz (2.2) can be reduced to the form (3.31) with
$U_{\rm in} ({\bf x})=1$, $U_{\rm out} ({\bf x})= U_{\rm f}({\bf x})$
and $B_\mu^{\rm (in)out}({\bf x},t)=B_\mu({\bf x},t)$.
We have seen already that the gauge transformations
$U_{\rm in} ({\bf x})=1$ and $U_{\rm out} ({\bf x})= U_{\rm f}({\bf x})$
are continuous functions of ${\bf x}$
which, as ${\bf x} \to \infty$,
approach direction-independent constants which satisfies the
requirements of Section 2.
But the gauge fields $B_\mu^{\rm in}({\bf x},t)$ and
$B_\mu^{\rm out}({\bf x},t)$
on the right hand side of eqs. (2.2)
were required to have essentially {\it finite} support
in the ${\bf x}$-space at any fixed time
$ t$ and  to vanish at any ${\bf x}$ as time goes respectively to
$T_{\rm min}$ or $T_{\rm max}$. The first requirement of the
essentially {\it finite} support
in the ${\bf x}$-space is easily satisfied which follows
from eqs. (3.2), (3.32)
and the fact that for classical solutions
$\psi$ and $1-\rho$ are well localized pulses at early and late times.
On the other hand, the second requirement that the $B_\mu({\bf x},t)$
fields should
vanish at any ${\bf x}$ as time goes respectively to
$T_{\rm min}$ or $T_{\rm max}$ is not satisfied since at these times
$$ B_\mu({\bf x},t) \sim a_\mu (r,t) \sim \epsilon^{\mu \nu}
\partial_\nu \psi(r,t) / \rho^2 (r,t) \ ,
\eqno(3.35)$$
which does not vanish since the $\psi$-pulses move {\it undistorted}
and do not tend to zero at large early or late times.
Here we differ
from the claim made in Ref. [10] that the amplitude of $\psi$-pulses
vanishes at early and late times. This claim of Ref. [10]
(which mistakenly quotes Ref. [8] for the justification of the claim)
contradicts to the arguments
stated earlier in the Section as well as to the arguments of Ref. [8] and
the Fig. 2.

In order to apply the formalism of Section 2 to the case of
classical fields
(3.31)-(3.34) in the background, the background (3.31)-(3.34)
should be modified
at the early past and the far future to switch off the interaction of
the gauge fields with the fermions. This will be done now by
switching off the
gauge invariant degrees of freedom, $\psi$ and $\rho^2-1 $,
of the background field
(3.31)-(3.34) at early times, $t: \ T_{\rm min} \leq t < t_*$,
and late times,
$t: \ t_* < t < T_{\rm max}$,
$$ \psi \to 0, \quad \rho^2-1 \to 0 \quad \forall t: \
\{t<T_{\rm i} < t_*\}
\cup \{t>T_{\rm f}> t_*\} \ .
\eqno(3.36)$$

The fermion number violation
which occurs in such modified classical backgrounds
is given by eq. (2.40),
$$\langle \hat{N}_{\rm f} - \hat{N}_{\rm i} \rangle =
 \nu [U_{\rm out}] - \nu [U_{\rm in}] = \nu [U_{\rm f}]=
\sum_{n}{\it degree}(\varphi(r_{*n},t_{*n})) \ ,
\eqno(3.37)$$
and is independent on the way of how the interaction
is switched off at early and late times and neither it depends on the times
$T_{\rm i}$ and $T_{\rm f}$ as far as their absolute values are much greater
than ${\rm max}_n|t_{*n}|$.

The procedure described by (3.36) corresponds to the situation
of interest where
an initial coherent gauge
field configuration was produced in the course of quantum
collision at some early
time, $T_{\rm i}$, and then evolved classically before
decaying into quantum radiation
at some late time, $T_{\rm f}$. In this work we are interested  in the
violation of the fermion number which occurred during the classical evolution
of the initial coherent state before it decayed. We assume here that there
was no fermion number violation before the coherent field
was created or after it decayed.

The fermion number
in our approach is violated only during the classical evolution of the
initial coherent configuration and not at the moment of its creation or decay.
Equation (3.37) establishes a selection rule for fermion
number violation in the
background of a classical solution in the spherical ansatz:
{\it the change of the numbers of fermions is equal to the sum
of the degrees of
$\varphi$ in each singular point}.
This is an integer by construction while the topological charge
$Q$ is not [7-8].

As an example we consider a special case of a
L\"uscher-Schechter solution with
 $\tau_0=1$ and $\varepsilon=1$ depicted on Figs. 1 and 2.
This solution has a single singular point with the degree of
$\varphi$ being
equal to unity. Thus, the violation of the fermion number in the presence
of this solution is one, while its topological charge is non-integer [7].

\goodbreak
\bigskip

\centerline{\titlefnt 4. Discussion}

In this work we considered fermion number violation in the background
of a pure $SU(2)$ gauge field in Minkowski space using the method of
N. Christ [9] reviewed in Section 2.
Then the method was applied for the case of classical solutions in
the spherical
ansatz in the background.
Fermion number violation in such backgrounds was considered
 also in the past in Refs. [8] and [10].
We will first compare our results to the interpretation of Ref. [8].
Naively applying the anomaly equation (1.2),
the net number of fermions produced was interpreted in Ref. [8] as to be
given by a topological charge $Q$ of the solution in the background.
Since $Q$ is non-integer in general, the violation of the fermion number being
equal to $Q$ was treated in Ref. [8] in a quantum average sense. That is,
in every experiment the violation is an integer, but averaging over
the experiments
one can obtain a non-integer result according to [8].
We no longer believe in this conclusion.  In the present approach we switch off
the gauge invariant degrees of freedom of the background at early and
late times
assuming that the classical configuration did not exist forever, but
was created at some early time and decayed into quantum radiation at some late
time. This allowed us to treat fermions as free in the early past and
late future.
We believe that the question of the fermion production is not well defined
for the classical background which existed forever, since the fermions
are never free
in this case and the particle interpretation is a conceptual difficulty
in this case.

Similarly to our work,
the Christ's approach [9] was also used in Ref. [10] to calculate
the violation of the fermion number in presence of the classical solutions
in the spherical ansatz.
We do not agree, however, with the method of Ref. [10]
which relied on an incorrect assumption
that the $\psi$ field was vanishing at early and late times and, thus,
the method
of Ref. [10] of handling the Christ's formalism
cannot be applied to the case of the classical solutions.

It is also rather instructive to compare our result with the result
of Ref. [15]
where the violation of the fermion number was studied in the gauge theory
in the Higgs phase.
It was shown there that the number of fermions produced is equal
to the change of the winding number of the Higgs field. In our approach
we do not
have a fundamental Higgs field, but a Higgs-like field $\chi$, eq. (3.7),
appears in the
spherical ansatz which is a Higgs field of the (1+1)-dimensional
Abelian Higgs
model, eq. (3.3). Our selection rule then implies that
the change of the numbers of fermions is equal to the change of
the winding number
of the Higgs-like field $\chi$ which looks somewhat parallel to the result
of Ref. [15]. Nevertheless, the  method of Ref. [15] relies on the existence
of a gap in the fermion spectrum and cannot be applied to our theory with
massless fermions.

But the method of Section 2 {\it can} be used for the case of the background
being a classical solution of the gauge theory in the Higgs phase.
Applying the Christ's approach to this case we would readily reproduce the
result of Ref. [15]. This was done in Ref. [16].
In fact, the case of the Higgs phase is easier than the case of the
pure gauge theory
since the classical solutions in the Higgs phase dissipate at early
and late times
because the gauge field  becomes massive.
 In the background of dissipating classical
solutions fermions do become free in the early past and the late future and
the gauge background does not have to be switched off.
In this case the violation of the fermion number is always integer
[15-16] and
does not have much to do with the topological charge $Q$ which in
this case is not even
well defined [15].
This means that one should be rather careful with a naive interpretation
of the anomaly equation. In the Christ's approach [9] discussed in Section 2
this difficulty is avoided by introducing   the
``fermion" charges of the radiating gauge fields $q^{\rm out}$ and
$q^{\rm in}$
 which have nothing to do with the actual number of fermions and
 have to be
subtracted from the right hand side of the anomaly equation, see eq. (2.39).

The necessary condition for fermion number violation in our approach
is the vanishing of the $\rho^2$ field.
For the perturbative solutions of Ref. [8] the field $\rho^2$
never vanishes and the barion number violation is zero.
(This point was already addressed in Ref. [10].) Thus, the non-zero
fermion number
violation cannot be achieved in the framework of perturbation
theory even in the model with an unbroken gauge group.
The non-perturbative solutions [7] of the Section 3 behave like $1/g$
with the energy, eq. (3.13), $E \sim 1/g^2$. This implies that even in
the case of
a QCD-like theory the classical field which causes chiral fermion number
violation
can be constructed only from the number of initial particles of order of
$1/ \alpha$. This suggests that there is a sphaleron-like configuration
which is the top of the barrier separating vacua with different
chiral fermion numbers
even in QCD. Of course, QCD is a scale-invariant model on the classical level
and the top of the barrier depends on an arbitrary scale which is supposingly
fixed by quantum effects.
In Section 3 we saw that the violation of the fermion number occurs when
the classical field passes through the ``singular" point where $\rho^2=0$.
For the specific case of L\"uscher-Schechter solutions fermion number is
violated by $\pm 1$
when the field $q$ of the associated mechanical problem
passes over the top, $U=1/2$, of the double well potential,
$U={1\over 2} q^2\hbox{(q+2)}^2$.
We associate the configuration (3.16) with $q=-1$,
$$ \rho^2 = \sin^2 w, \quad \psi=0,
\eqno(4.1)$$
with a sphaleron-like configuration in QCD. The configuration (4.1)
is, in fact, a classical (time-dependent) solution of de Alfaro,
Fubini and Furlan [17].
The interpretation of this solution as an exploding sphaleron of
QCD was recently
made in Ref. [18].

If the quantum effects of QCD fix the scale, then some quantum analog
of de Alfaro-Fubini-Furlan solution will become a real quantum
sphaleron of QCD
and for the energy below the quantum sphaleron mass the chirality
violation
for massless fermions will not occur, while for energies higher than this
mass the violation may very well happen.

\vskip 1truecm
\centerline{\bf Acknowledgments}
I am grateful to Eddie Farhi and Bob Singleton for infinite discussions
about classical solutions.
I would like to thank N. Christ, J. Goldstone and N. Manton for
discussions on anomalous fermion number violation.
I also benefited from T. Gould, S. Hsu and K. Rajagopal and from early
discussions with V.A. Rubakov and L. Yaffe on fermion number violation.

\vfill
\eject

\centerline{\bf REFERENCES}
\medskip
\item{[1]}  G. 't Hooft,  {\it Phys. Rev.} {\bf D14} (1976) 3432;
(E) {\bf D18}
  (1978) 2199.
\item{[2]}  S. Adler, {\it Phys. Rev.} {\bf 177} (1969) 2426;
 \item{} J. Bell and R. Jackiw, {\it Nuovo Cimento} {\bf 51} (1969) 47.
\item{[3]}  N. Manton, {\it Phys. Rev.} {\bf D28} (1983) 2019;
 \item{} F. Klinkhamer and  N. Manton, {\it Phys. Rev.} {\bf D30} (1984)
 2212.
\item{[4]}  A. Belavin, A. Polyakov, A. Schwarz and  Yu. Tyupkin,
  {\it Phys. Lett.} {\bf B59} (1975) 85.
\item{[5]}  A. Ringwald, {\it Nucl. Phys.} {\bf B330} (1990) 1.
\item{[6]} M. Mattis, {\it Phys. Rep.} {\bf 214} (1992) 159;
\item{}P.G. Tinyakov, {\it Int. J. Mod. Phys.} {\bf A8} (1993) 1823.
\item{[7]}
E. Farhi, V.V. Khoze and R. Singleton, {\it Phys. Rev.} {\bf D 47}, 5551
(1993).
\item{[8]} E. Farhi, V.V. Khoze, K. Rajagopal and R. Singleton,
{\it Phys. Rev.} {\bf D 50}, 4162 (1994).
\item{[9]}  N. Christ, {\it Phys. Rev.} {\bf D21} (1980) 1591.
\item{[10]} T.M. Gould and S.D.H. Hsu, Harvard preprint HUTP-94/A036,
hep-ph/9410407.
\item{[11]} R. Jackiw and K. Johnson, {\it Phys. Rev.} {\bf 182} (1969) 1459.
\item{[12]} E. Witten,    {\it Phys. Rev. Lett.} {\bf 38} (1977) 121;
 \item{} B. Ratra and L.G. Yaffe, {\it Phys. Lett.} {\bf B205}  (1988) 57.
 \item{[13]} M. L\"uscher, {\it Phys. Lett.} {\bf B70}   (1977) 321;
 \item{} B. Schechter, {\it Phys. Rev.} {\bf D16} (1977) 3015.
\item{[14]} S. Wolfram, {\it Mathematica, A System for Doing Mathematics by
Computer}, $2^{\rm nd}$ Edition, Addison Wesley 1992.
\item{[15]} E. Farhi, J. Goldstone, S. Gutmann, K. Rajagopal and R. Singleton,
MIT preprint CTP 2370, hep-ph/9410365.
\item{[16]} T.M. Gould and S.D.H. Hsu, Harvard preprint HUTP-94/A039,
hep-ph/9410408.
\item{[17]} V. de Alfaro, S. Fubini and G. Furlan,
{\it Phys. Lett.} {\bf 65B}   (1976) 163.
\item{[18]} G. Gibbons and A. Steif, Cambridge preprint DAMTP/R94/53.
\vfil\eject

\centerline{ \bf Figure Captions}
\medskip
\item{\bf Fig. 1} The $\rho^2$-component of
a L\"uscher-Schechter solution with $\varepsilon=1$ and $\tau_0=1$.
Figure 1a shows the (incoming) $r$-profiles of $\rho^2$ for a sequence
of negative times: $-10<t<0$. Figure 1b shows the (outgoing) $r$-profiles
of $\rho^2$ for a sequence
of positive times: $0<t<10$.
\medskip
\item{\bf Fig. 2} The $\psi$-component of
a L\"uscher-Schechter solution with $\varepsilon=1$ and $\tau_0=1$.
Figure 2a shows the (incoming) $r$-profiles of $\psi$ for a sequence
of negative times: $-10<t<0$. Figure 2b shows the (outgoing) $r$-profiles
of $\psi$ for a sequence
of positive times: $0<t<10$.
\medskip
\item{\bf Fig. 3} Contours of integration,
${\cal C}_{(r,T_{\rm min})\mapsto (r,t)}$, used in eq. (3.30)
to define a continuous gauge transformation are shown
for two cases: 1) $r\equiv r_1 < r_*$; and 2)$r\equiv r_2 > r_*$.

\vfill
\eject

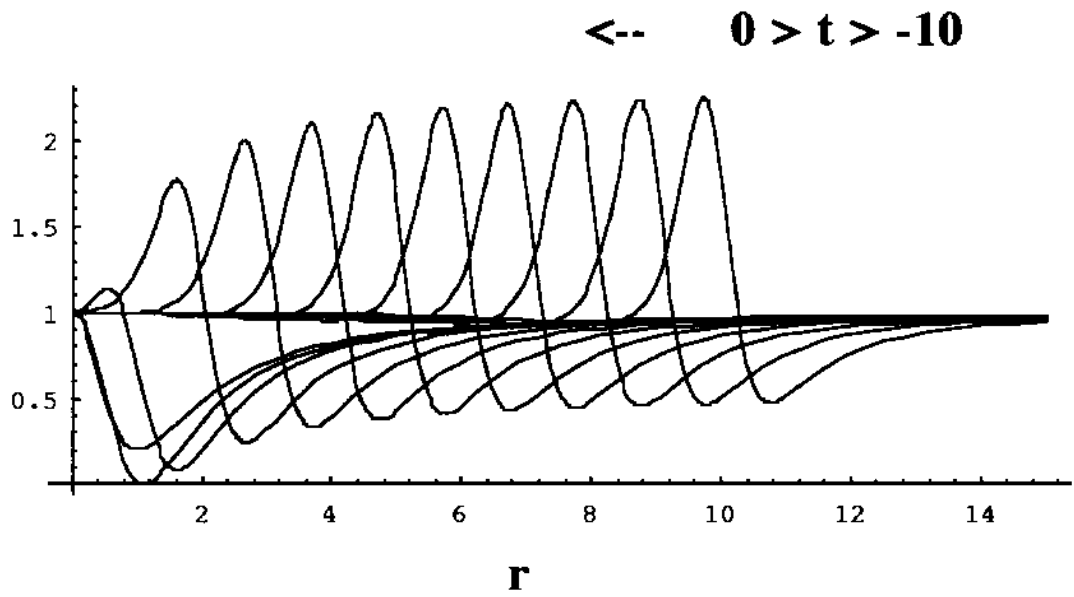
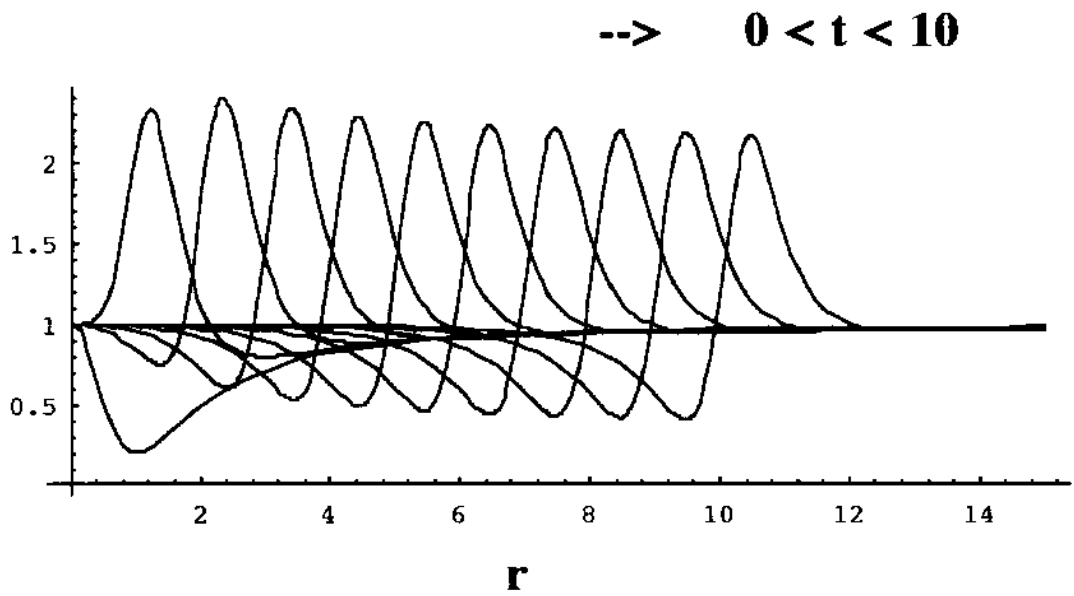
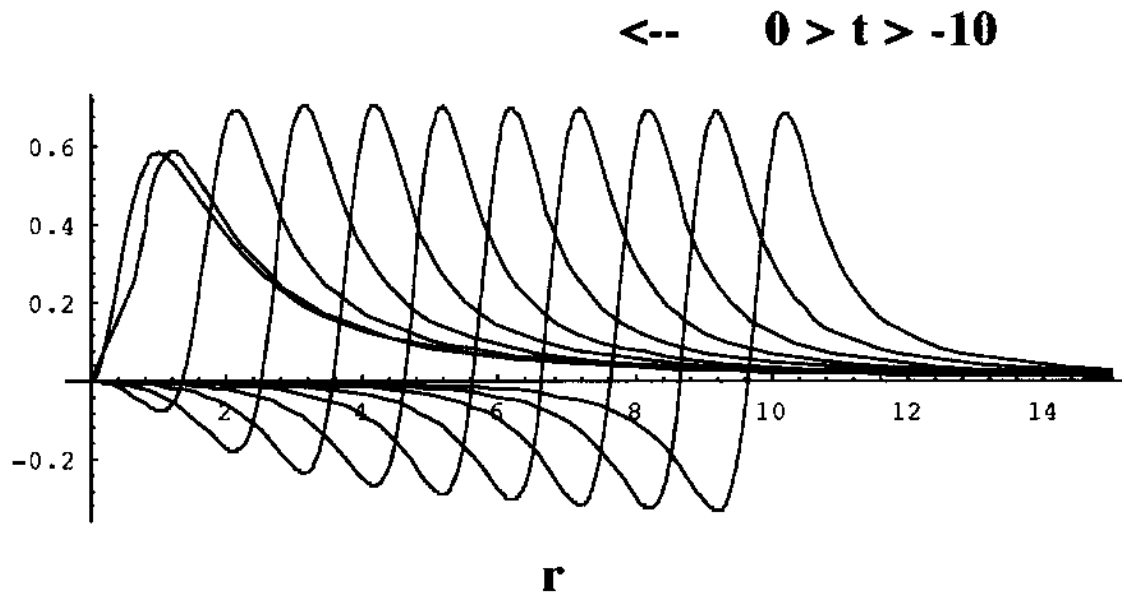
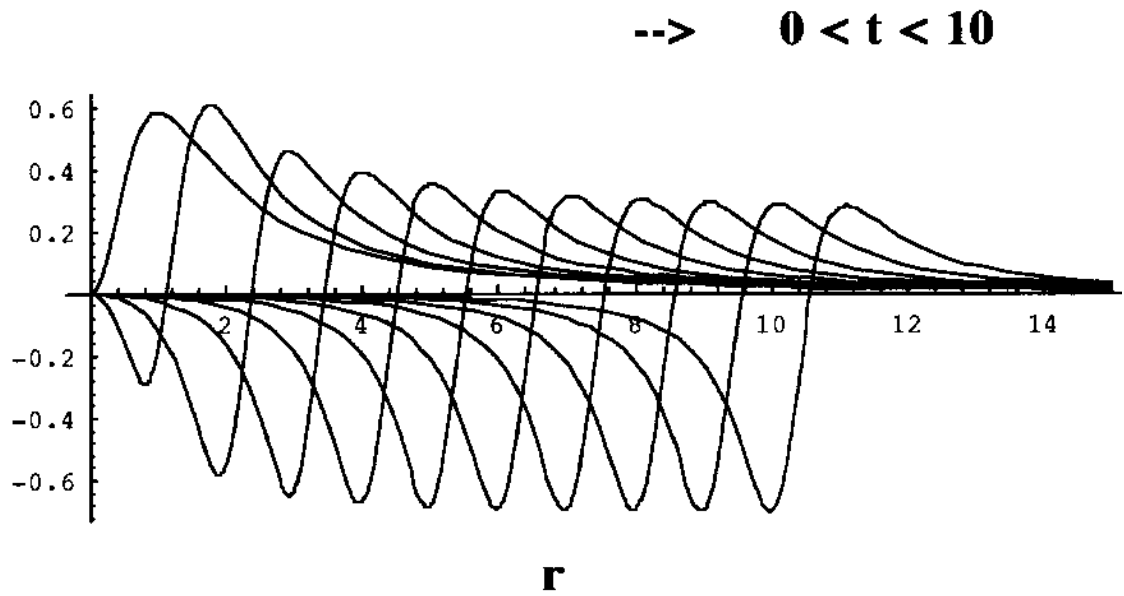
\bye